\documentstyle[aps]{revtex}
\input psfig.tex
\begin{document}
\preprint{\parbox[b]{3.3cm}{NPL-1138}}
%\draft
%%%%%%%%%%%%%%%%%%%%%%%%%%%%%%%%%%%%%%%%%%%%%%%%
\begin{title}
{Evolution of Gluon Spin in the Nucleon}
\end{title}
\author{S.J.~Pollock}
\address{University of Colorado, Boulder CO, USA 80303
\\Email address: pollock@lucky.colorado.edu
}
\date{\today}
\maketitle
\begin{abstract}

We examine the $Q^2$ evolution of gluon polarization in polarized
nucleons.  As is well known, the evolution of $\alpha_s \Delta G(Q^2)$
is negligible for typical momentum transfer variations found in
experimental deep inelastic scattering.  As $\alpha_s$ increases,
however, the leading nonzero term in the evolution equation for the
singlet first moment reduces the magnitude of the gluon spin.  At low
$Q^2$ the term $\alpha_s \Delta G$ can vanish, and ultimately become
negative.  Thus, low energy model calculations yielding negative
$\Delta G$ are not necessarily in conflict with experimental evidence
for positive gluon polarization at high $Q^2$.

\end{abstract}
\bigskip
\newpage
%%%%%%%%%%%%%%%%%%%%%%%%%%%%%%%%%%%%%%%%%%%%%%%%%%%

\section{Introduction}

Polarized deep inelastic scattering (DIS) has proven to be an important
new tool for studying the structure of the nucleon.  Through sum rules,
the experiments have provided us with values of various operator matrix
elements; for instance, the singlet sum rule indicates that quark spin
accounts for only a small fraction ($\approx$ 1/3) of the nucleon's
total spin\cite{oldexpt,smc97}.  One possible resolution of
this apparently surprising result is that the gluon fields in the
nucleon bound state may carry a significant fraction of the nucleon's
spin. Orbital angular momentum of quarks and gluons may also be
significant.

Recently, various authors\cite{jaffe95,glue} have
attempted to make estimates of the size of the gluon contribution to
the spin of the nucleon, 
\begin{equation}
\Delta G(Q^2) = \int_0^1 dx \Delta g(x,Q^2) =
\int_0^1 dx(g_\uparrow(x,Q^2)-g_\downarrow(x,Q^2)).
\end{equation}
In at least one case\cite{jaffe95}, this quantity was evaluated in low
energy models, and turned out to be {\it negative}. This is in apparent
conflict with extractions, primarily at DIS scales, which find $\Delta
G$ positive\cite{smc97,ab2}. Indeed, to leading order (LO) in perturbative
QCD, $\alpha_s(Q^2) \Delta G(Q^2)$ is renormalization group invariant,
thus the claim has been made that at least the sign of $\Delta G$
should be a reliable prediction of the models\cite{jaffe95}.

It is well known\cite{history} that the singlet part of the first
moment of the spin structure function $g_1$ has an anomalous $Q^2$
evolution. It is also well known\cite{kod80,larin94} that the {\em leading}
term in $\alpha_s$ in the axial anomalous dimension vanishes, and it is
for this reason that many authors dismiss this $Q^2$ evolution as
insignificant.  Roughly speaking, corrections to the singlet first
moment arising from the anomalous dimension can be argued to behave
like $\alpha_s \log Q^2$, and hence appear {\em approximately} $Q^2$
independent. This is, however, not precisely true at higher order in
$\alpha_s$.  In an earlier paper,\cite{kunz89} it was argued that
comparing momentum sum rules from unpolarized electroproduction and the
spin sum rule for $g_1$, including their $Q^2$ evolution, showed that
DIS spin measurements are {\em consistent} with a low energy valence
quark picture, where the valence quarks carry a substantial part of the
spin of the proton. In a later paper,\cite{mulders95}, those
calculations were extended to consistently include the next higher
order QCD corrections in leading twist.  Here, we will explicitly
examine the effects on gluon spin.  We evaluate the leading non-zero
QCD corrections to the gluon spin evolution, and find that the sign of
the first correction term forces one in the direction of smaller (more
negative) $\alpha_s \Delta G$ at lower $Q^2$.  Application of these
perturbative formulae show that a sign change in $\alpha_s \Delta G$
thus need not be such a surprise.

There are several caveats to this analysis which must be made at the
outset. First, we apply perturbative formulae down to a regime where
they cannot rigorously apply ($\alpha_s/\pi\approx 1$) and thus do not
expect to have quantitative predictive power. However, the sign change
in $\Delta G$ is significant, and if the perturbative result goes in
this direction, there is certainly no reason why the full
nonperturbative result might not behave similarly. Furthermore, one
can point to the poorly understood successes of the one-gluon exchange
mechanism in the hadron mass spectrum\cite{rujula75}, or of the
evolution of unpolarized structure functions\cite{kunz89}, which
provide some faith in at least the qualitative predictions of this
method. 

The second issue is one of scheme dependence. The decomposition of the
singlet sum into terms involving $\Delta \Sigma$ and $\Delta G(Q^2)$
depends on renormalization and factorization scheme choice, and is
gauge dependent\cite{schemes}. Indeed, there is no unique way to
define quark and gluon distributions beyond LO.  When in a
perturbative QCD regime, we work in the Adler-Bardeen (AB)
factorization scheme\cite{ab2,larin94,mulders95,ab69}, with
$\overline{MS}$ renormalization conventions, which defines $\Delta G$
and its contribution to the singlet first moment as well as its $Q^2$
evolution to LO.  However, if one wants to work beyond
leading order,  in particular to evolve to low scales to compare with
model predictions, one must know in addition the connection with the
choice of scheme (and thus the underlying operator matrix element
definition) of e.g. $\Delta G$ in the low energy model calculations.
Thus, we really cannot make any direct quantitative comparisons with
the numerical values extracted in quark model calculations. We can,
nevertheless, unambiguously examine the leading evolution of the gluon
spin downwards in Q$^2$ given our choice of scheme,  which shows a
clear trend towards zero and eventually negative values.  Ultimately,
one must understand both the renormalization scale and scheme
dependence of the quark model calculations in order to decide if the
contribution of gluon spin is quantitatively compatible at low and high
energy scales.

% In the following section we introduce some of the formalism required,
%% indicate the evolution equations and connections to observables.  In
%% section three, we present our results and conclusions.

\section{Formalism}

Working at leading twist, the contributions to the hadronic tensors
involved in polarized DIS, via the operator product expansion, are due
to singlet and non-singlet axial currents. The first moment of the
polarized spin structure function $g_1(x,Q^2)$ is given by
\begin{equation}
\Gamma_1^{p(n)} = \int_0^1 dx g_1^{p(n)} = {1\over9}C^S(Q^2)
a_0(Q^2) + {1\over6}C^{NS}(Q^2)(\pm a_3/2 + a_8/6) 
\label{eqa}
\end{equation}
where $C^S(NS)$ are (non) singlet coefficient functions 
calculable in perturbative QCD which have been worked out to
second(third) order in $\alpha_s$\cite{smc97,kod80,larin94}. Both are
given by $1-\alpha_s(Q^2)/\pi$ to leading order. $a_3$ and $a_8$ are
renormalization group invariants, and can be obtained (modulo 
issues of flavor SU(3) symmetry breaking\cite{su3}) from low energy
$\beta$-decay constants.

The singlet matrix element is obtained from
\begin{equation}
a_0(Q^2)s_\mu = 
\langle p,s|J_\mu^5|p,s\rangle = 
\langle p,s|\sum_{i=1}^{n_f}\bar q_i \gamma_\mu\gamma_5 q_i |p,s\rangle,
\end{equation}
which in a naive quark model picture is just the total spin of
the quarks.  Experimental results are generally given for values of
$a_0(Q^2)$ with the coefficient function $C^S(Q^2)$ factored out, as
shown above, but because of the U(1) anomaly in QCD, $a_0$ is itself
not scale independent.  The QCD evolution of $a_0$ has been worked out
to NNLO\cite{larin94}, and is given below.  For our purposes, we also
wish to consider a decomposition of $a_0$ into separated quark and
gluon contributions. This {\it is} a renormalization scheme-dependent,
and gauge dependent, separation. Some authors {\it e.g.} choose
a so called gauge-invariant scheme in which the gluons by definition do
not contribute to $a_0$\cite{schemes}. In the Adler-Bardeen factorization
scheme\cite{ab2,ab69}, sometimes called a chirally invariant scheme,
the anomalous gluon contribution is explicitly separated out as
follows:
\begin{equation}
a_0(Q^2) = \Delta \Sigma - f \alpha_s(Q^2) \Delta G(Q^2)/2 \pi
\label{eqa0}
\end{equation}
where f is the number of active flavors at this $Q^2$. 
The evolution of the full singlet first moment, $a_0$, is known to next
to leading order (NLO) in $\alpha_s$ \cite{larin94,mulders95},
\begin{equation}
a_0(Q^2)=
a_0(Q_0^2)\,
\exp\left(-\int_{\alpha_s(Q_0^2)}^{\alpha_s(Q^2)}
	d\alpha'
	{\gamma^S(\alpha')\over 2\beta(\alpha')}\right)\label{a0ev}
\end{equation}
where
%\begin{equation}
%\gamma^S(\alpha_s)= \gamma_1^S\left(\alpha_s/4\pi\right)^2+
%\gamma_2^S \left(\alpha_s/4\pi \right)^3\cdots ,
%\end{equation}
$\gamma^S(\alpha_s)= \gamma_1^S\left(\alpha_s/4\pi\right)^2+
\gamma_2^S \left(\alpha_s/4\pi \right)^3 + \cdots $
is the singlet anomalous dimension, with $\gamma_1^S=16f$, and
$\gamma_2^S= (944 f/3- 32 f^2/9)$ with our choice of renormalization
conventions, and $\beta$ is the beta function of QCD,
\begin{equation}
\beta(\alpha_s)\ = \ d\alpha_s/d\tau \ 
=\ -\beta_0 {\alpha^2\over 4 \pi} - 
		\beta_1 {\alpha^3\over 16 \pi^2}+\ldots
\label{alphas}
\end{equation}
where $\tau = \ln(Q^2/\Lambda_{QCD}^2)$,  and 
$\beta_0=11-2f/3$, $\beta_1=102-38f/3$ \cite{BETA}.
To leading non-zero order, this gives 
\begin{equation}
a_0(Q^2) \approx a_0(Q_0^2)
\left(1+{\gamma_1^S\over 8\pi \beta_0}
	\left(\alpha_s(Q^2)-\alpha_s(Q_0^2)\right)+ 
			{\cal O}(\alpha_s^3)\right)\label{aeqn}
\end{equation}
Note that the {\it difference} $\alpha_s(Q^2)-\alpha_s(Q_0^2)$
appearing in the formulae above  is itself of order $\alpha_s^2$, thus
naturally the evolution of $a_0$ vanishes at order $\alpha_s$.
The solution for $a_0$ including NLO corrections in the singlet
anomalous dimension and the beta function is \cite{larin94,mulders95}
\begin{eqnarray}
a_0(Q^2)&=&a_0(Q_0^2)
\biggl( 1 + \frac{\gamma_1^S}{8\pi \beta_0}
(\alpha_s(Q^2)-\alpha_s(Q_0^2))
+ \left(\frac{\beta_0 \gamma_2^S - \beta_1 \gamma_1^S}{64 \pi^2 \beta_0^2}
                \right) (\alpha_s^2(Q^2)-\alpha_s^2(Q_0^2))
\nonumber \\
&&\qquad\qquad\qquad\qquad\qquad
+ \frac{(\gamma_1^S)^2}{128 \pi^2 \beta_0^2}
 (\alpha_s(Q^2)-\alpha_s(Q_0^2))^2
+{\cal O}(\alpha_s^4)\biggr), \label{aeqnlo}
\end{eqnarray}
truncated at order $\alpha_s^3$. 

It is important to emphasize that the evolution of $a_0(Q^2)$ is well
defined, and is not itself gauge dependent.  It is the {\it separation}
of $a_0$ into ``spin" and ``gluon" terms which is scheme and gauge
dependent.  
%Could skip here to next comment%
Using the separation choice of Eq.~(\ref{eqa0}), the
evolution of these quantities are given
by\cite{larin94,kunz89,mulders95}
\begin{equation}
{d\over d\tau}\left(\Delta\Sigma(\tau)\atop \alpha_s(\tau) \Delta
G(\tau)/2\pi\right) =
\alpha_s^2(\tau)\left(\matrix{
	\gamma_{\Sigma\Sigma}&\gamma_{\Sigma\Gamma}\cr
	\gamma_{\Gamma\Sigma}&\gamma_{\Gamma\Gamma}\cr}\right)
	\left(\Delta\Sigma(\tau)\atop 
		\alpha_s(\tau)\Delta G(\tau)/2\pi\right)\label{eqm}
\end{equation}
To leading order,
$\gamma_{\Gamma\Sigma}=1/2\pi^2$, 
$\gamma_{\Sigma\Sigma}=0$,
$\gamma_{\Sigma\Gamma}=0$, 
$\gamma_{\Gamma\Gamma}=-f/2\pi^2$
\cite{kunz89}.
The evolution manifestly begins only at second order in $\alpha_s$.
% Could skip above, would have to rewrite the following.%
To leading non-zero order, solving Eq.~(\ref{eqm}) directly (or
alternatively, combining Eqs.~\ref{eqa0} with \ref{aeqn}) we see
\begin{equation}
\alpha_s(Q^2)\Delta G(Q^2) = 
\alpha_s(Q_0^2)\Delta G(Q_0^2)  -{\gamma_1^S\over 4 f \beta_0}
\left(\alpha_s(Q^2)-\alpha_s(Q_0^2) \right)\,a_0(Q^2).
\label{adg}
\end{equation}
Of course, to {\it first order} in $\alpha_s$, the above equation
states that $\alpha_s \Delta G$ is a renormalization group invariant,
the usual result.  This last equation already shows qualitatively the
results we have claimed; namely, if $Q^2$ is some large scale
appropriate to DIS, and $Q_0^2$ is some low scale appropriate to a
quark model calculation, the leading nonzero correction to $\alpha_s
\Delta G$ is large and positive, and there is no reason the sign need
be preserved between low and high $Q^2$ scales.  In the Adler-Bardeen
scheme, Eq.~(\ref{eqa0}) does not receive higher order
corrections\cite{ab2}, and one can combine Eq.~(\ref{eqa0}) with
Eq.~(\ref{aeqnlo}) to examine the NLO modifications to $\alpha_s \Delta
G$.

\section{Results}

For definiteness, we use the following experimental
values\cite{smc97}:  $\alpha_s(5 {\rm\ GeV}^2)=.287\pm .02$,
$a_0(5{\rm\ GeV}^2) = 0.37\pm .11$.  The evolution equation,
(\ref{a0ev}) evaluated to LO (NLO) yields 
\begin{equation} {a_0(5 {\rm\ GeV}^2)/
a_0(10 {\rm\ GeV}^2)  }= 1.008 (1.010), 
\end{equation} 
which is a negligible change, well below the present limits of
experimental observability. In this sense, the fact that to leading
order there is no $Q^2$ evolution is born out. Leaving the region of
small $\alpha_s$, however, yields some nontrivial evolution.  As
discussed above,  the separation of $a_0$ into $\Delta \Sigma$ and
$\Delta G$ requires additional experimental and theoretical
assumptions; recent analyses \cite{smc97,ab2} consistent with the
scheme conventions of this work finds $\Delta G(5 {\rm\ GeV}^2)$
ranging from 0.8 to 2.6, and thus $\alpha_s \Delta G(5\ {\rm GeV}^2)$
ranging from 0.2 to 0.75, and $\Delta \Sigma$ (which is scale {\it
independent} here) ranging from 0.4 to 0.65, depending on how the
analysis and extraction are done. The most naive predictions originally
gave $\Delta \Sigma=1$, but many effective low energy quark models are
consistent with a lower value, around $\Delta \Sigma \approx
0.65$\cite{kunz89}. (The reduction from unity might come {\it e.g.}
from the lower components of the relativistic quark spinors, the same
source which reduces the axial charge in the bag model from 5/3) Using
any such prediction in Eq.~(\ref{eqa0}), however, {\it assumes a
partonic interpretation} of $\Delta\Sigma$, which is by no means
required by QCD, and may be entirely incorrect\cite{interp}.

Starting from the above experimental values, $\alpha_s \Delta G$ will
evolve from somewhere in the range $(.2\leftrightarrow.75)$ at 5
GeV$^2$ down to zero at a scale where $\alpha_s/\pi$ is in the range
$(.3 \leftrightarrow 1.5)$, {\it i.e.} where nonperturbative physics
should just begin to become significant. Going further down in $Q^2$,
$\alpha_s$ increases and $\alpha_s\Delta G$ becomes negative.  A
smaller assumed value for $a_0(5{\rm\ GeV}^2)$ results in {\it less}
rapid downward evolution.  In Figs.  1-3, we show the result of
evolving the singlet moment as well as the gluon spin up in $\alpha_s$
(down in $Q^2$).  If one applies the NLO corrections of
Eq.~(\ref{aeqnlo}) to $a_0$, this makes the evolution slightly {\it
more} rapid. That is, one need not go so far in $\alpha_s/\pi$ to get
to the ``turnover" point ($\Delta G=0$, or $a_0\approx 0.5$)
Explicitly, $a_0$ reaches 0.5 at $\alpha_s\approx 2$ using LO evolution
for $a_0$, but $\alpha_s\approx 1.3$ using NLO.  It is encouraging that
the trend to smaller (more negative) $\Delta G$ at low $Q^2$ is
enhanced (rather than cancelled, as it certainly might have been) by
considering the NLO corrections of Eq.~(\ref{aeqnlo}).

Moving the other direction, one can ask how far the trend of increasing
$\alpha_s\Delta G$ with increasing $Q^2$ continues. From
Eq.~(\ref{aeqn}), as $Q^2$ increases, $a_0(Q^2)$ (slowly) decreases and
this in turn {\it reduces} the rate of change of $\alpha_s \Delta G$.
In this way, $\alpha_s\Delta G$ inevitably increases, but assuming that
$\Delta\Sigma$ is in the range  $(.5 \leftrightarrow 1)$, Eq.~(\ref{adg})
shows that if $\alpha_s \Delta G$ reaches around $(1\leftrightarrow
2)$, the increase with scale effectively halts, and $\alpha \Delta G$
becomes renormalization group invariant {\it beyond} second order in
$\alpha_s$, as well.

Following the logic of refs \cite{kunz89} and \cite{mulders95},
we might argue that evolution of both unpolarized momentum fractions
and polarized spin fractions are consistent with a nonrelativistic
quark picture at a low energy scale of $\alpha_s(\mu_0^2)/\pi\approx
1\pm 0.2$, where the gluonic contributions vanish. Thus, with $\Delta
G(\mu_0^2)=0$, and $\Delta\Sigma=.5$\cite{mulders95}, we would predict,
to leading non-zero order,
\begin{equation}
\alpha_s\Delta G(5 {\rm\ GeV}^2) = 0-
(4/ \beta_0)
(\alpha_s(5\ {\rm GeV}^2)-\alpha_s(\mu_0^2))(\Delta\Sigma) \approx  
0.6\pm .1,
\end{equation}
{\it i.e.}
$\Delta G(5\ {\rm GeV}^2)\approx 2.1 \pm .4$,
a result which is certainly compatible with recent high energy
experimental estimates\cite{smc97}, even though we begin with $\Delta
G=0$ at the low energy scale.

Jaffe provides us with more sophisticated model predictions for $\Delta
G$ at the low energy scale\cite{jaffe95}, namely for the
non-relativistic quark model $\Delta G({\rm NRQM})=-0.8$ and
$\alpha_s({\rm NRQM})=0.9$, while for the bag model $\Delta G({\rm
bag}) = -0.2$, and $\alpha_s({\rm bag})=2.0$.  His paper makes it clear
that these numbers are not to be taken {\it too} seriously, but the
sign of $\Delta G$ is taken as an interesting and surprising result.
However, if we evolve these as above (bearing in mind the caveats
outlined in the introduction) we would predict $\alpha_s \Delta
G(5\ {\rm GeV}^2)({\rm from\ NRQM}) = -.5$, while $\alpha_s \Delta
G(5\ {\rm GeV}^2)({\rm from\ bag}) = +.1$. The latter has shown a sign
change, with $\Delta G(5\ {\rm GeV}^2)=+.35$ only slightly below the
low end of the experimental range. Assuming a (scale independent) value
of $\Delta\Sigma=.7$, rather than .5, brings these predictions even
higher, e.g. from the bag model result it predicts $\Delta G(5\ {\rm
GeV}^2) = 1$. We see that the sign can indeed change, although it
depends on both the starting values for $\alpha_s$ and $\Delta G$, and
thus on the quantitative details of the model.  This is also clear from
the evolution shown in Figs. 2 and 3, where the location (in
$\alpha_s$) of the sign change is a strong function of the starting
values.

We conclude that the sign of $\alpha_s \Delta G$ at {\it low energy}
scales is ill-determined simply based on ${\cal O}(\alpha_s^2)$ (which
is leading non-vanishing order) perturbative QCD arguments, starting
from experimental values at moderate $Q^2$.  Both LO and NLO
perturbative evolution tend to decrease $\alpha_s \Delta G$, even to
negative values, depending on starting conditions.  The spin content of
the glue remains an important aspect of nucleon structure, one whose
measurements should certainly be improved at high $Q^2$, and which may
play a key role in understanding and interpreting low energy models,
but the {\it connection} between the two scales involves
nonperturbative physics and is thus not trivial to predict, and one
should not be particularly surprised if $\alpha \Delta G$ is found to
be negative in a quark model calculation.

\acknowledgments

The author acknowledges useful discussions with E.
Kinney.  This work is supported by U.S.  Department of Energy grant
DOE-DE-FG0393ER-40774. The author acknowledges the support of a
Sloan Foundation Fellowship.

\newpage
\begin{figure}
\vskip .2in
\psfig{figure=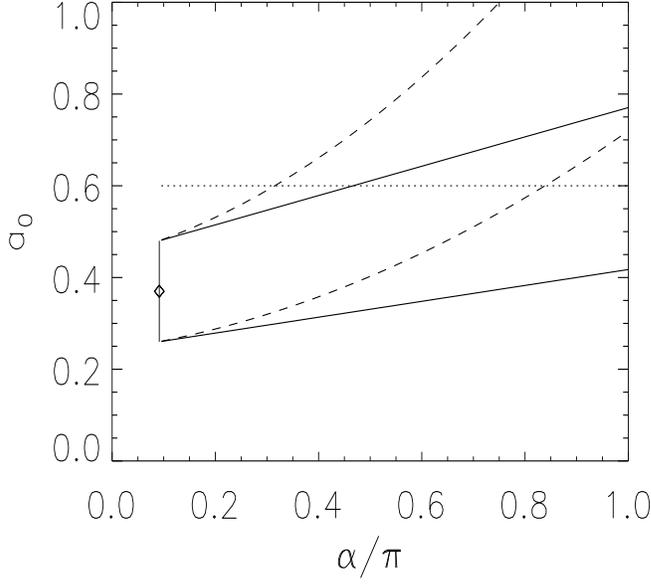,height=3.6in,width=4.0in}
\caption{Plot of the singlet first moment $a_0(Q^2)$, as a function of
$\alpha_s/\pi$. Solid (dashed) curve shows leading (next to leading) order
calculations. We started from experimental values\protect\cite{smc97}
at 5 GeV$^2$ and evolve downwards in $Q^2$.  The two lines represent a
band of possible results, beginning at $a_0(Q^2=5\ {\rm GeV}^2) = .37
\pm .11$.  The horizontal dotted line shows $a_0 = 0.6$, which would be
the rough expectation in the naive quark model picture, {\it assuming}
that $\Delta \Sigma$ (scale independent) corresponds to the ``spin on
the quarks", when there is no significant contribution from the gluon
anomaly term in Eq.~(\protect\ref{eqa0}).  }
\label{figi} \end{figure}

\begin{figure}
%\vskip .2in
\vskip .2in
\psfig{figure=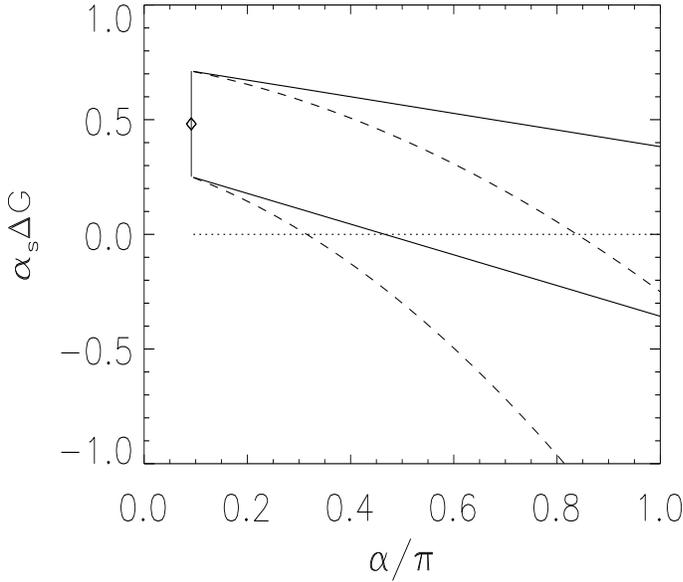,height=3.6in,width=4.0in}
\caption{Same as previous figure, but showing $\alpha_s \Delta G$, as
calculated in Eq.~(\protect\ref{adg}), as a function of $\alpha_s/\pi$.
The dashed curve corresponds to using $a_0$ as computed in
Eq.~(\protect\ref{aeqnlo}), to NLO. The horizontal dotted line is $\alpha_s
\Delta G=0$, where the gluon anomaly term makes no contribution to
Eq.~(\protect\ref{eqa0}).}
\label{figii} 
\end{figure}

\begin{figure}
%\vskip .2in
\vskip .2in
\psfig{figure=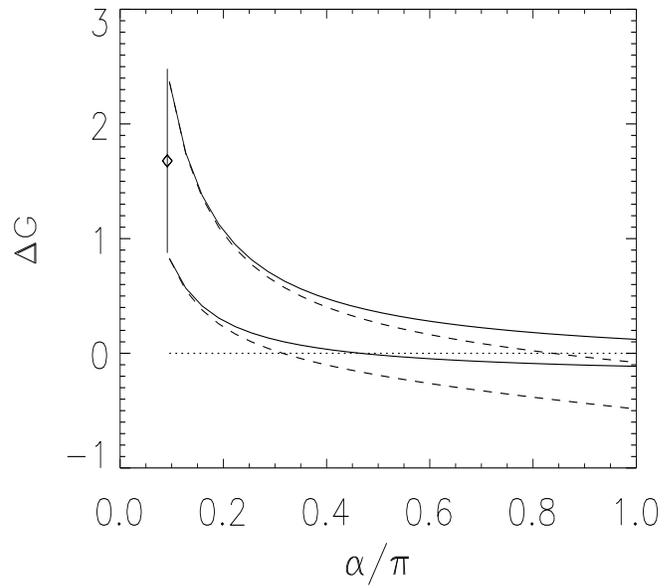,height=3.6in,width=4.0in}
\caption{Same as previous figure, but showing just $\Delta G$
as a function of $\alpha_s/\pi$, evaluated using Eq.~(\protect\ref{adg}).
The dashed curve corresponds to using $a_0$ as computed in
Eq.~\protect\ref{aeqnlo}, to NLO.}
\label{figiii} 
\end{figure}

\end{document}